# Hedging Effectiveness under Conditions of Asymmetry

## Abstract

We examine whether hedging effectiveness is affected by asymmetry in the return distribution by applying tail specific metrics to compare the hedging effectiveness of short and long hedgers using crude oil futures contracts. The metrics used include Lower Partial Moments (LPM), Value at Risk (VaR) and Conditional Value at Risk (CVAR). Comparisons are applied to a number of hedging strategies including OLS and both Symmetric and Asymmetric GARCH models. Our findings show that asymmetry reduces in-sample hedging performance and that there are significant differences in hedging performance between short and long hedgers. Thus, tail specific performance metrics should be applied in evaluating hedging effectiveness. We also find that the Ordinary Least Squares (OLS) model provides consistently good performance across different measures of hedging effectiveness and estimation methods irrespective of the characteristics of the underlying distribution.

John Cotter, Director of Centre for Financial Markets, Department of Banking and Finance, University College Dublin, Blackrock, Co. Dublin, Ireland, tel 353 1 716 8900, e-mail john.cotter@ucd.ie. Jim Hanly, School of Accounting and Finance, Dublin Institute of Technology, tel 353 1 402 3180, e-mail james.hanly@dit.ie. The authors would like to thank the participants at the Global Finance Annual Conference for their constructive comments.

1.    Introduction

A large literature has developed in the evaluation of futures based hedging strategies. The dominant hedging framework uses the variance as the risk measure and the Optimal Hedge Ratio (OHR) has, therefore, become synonymous with the minimum variance hedge ratio (see, for example Lien and Tse, 2002, for a comprehensive survey). A shortcoming of the variance risk measure is that it cannot distinguish between positive and negative returns and therefore it does not provide an accurate measure of risk for asymmetric distributions. Given asymmetry, hedging effectiveness metrics that cannot distinguish between tail probabilities may be inaccurate in terms of risk measurement.

In the literature on optimal hedging, two broad strands have emerged that address these issues. The first approach is the use of hedging estimation methods that seek to minimise some measure of risk other than the variance. These include LPM and methods relating to VaR and CVaR. The second approach is the use of hedging estimation methods that allow for asymmetry in the return distribution. Of these, asymmetric extensions to the GARCH class of models have been applied in a number of studies (see Brooks et al, 2002; and Lien, 2005a). However no study has addressed the issue of hedging effectiveness under conditions of asymmetry by combining these models, together with hedging effectiveness methods that measure tail probabilities. This study addresses this, and differs from prior research in a number of ways.

Firstly, we apply tail specific hedging effectiveness measures together with estimation methods that allow for asymmetry to crude oil spot and futures contracts. This allows us to compare the hedging effectiveness of short and long hedgers under conditions of asymmetry. Oil commodity markets were chosen as they are particularly suited for an examination into the effects of asymmetry on hedging effectiveness given their distributional characteristics and in particular, their tendency to be non-symmetrically distributed (see, for example, Baillie and Myers, 1991, Lien and Wilson, 2001). The cash prices of crude oil tend toward excess skewness given the volatile nature of the market and the underlying supply conditions. Also, since the futures price of a commodity is driven by storage costs and the convenience yield, the relationship between them may give rise to asymmetric returns. Secondly, we use empirical data based on distributional characteristics whereby a symmetric and an asymmetric dataset were chosen specifically so as to compare hedging effectiveness for both normal and asymmetric distributions.

The hedging estimation models used are; naïve hedge, rolling window OLS hedge, and two bivariate GARCH models including an asymmetric GARCH model. Four different hedging effectiveness metrics are applied. These are based on; Variance, LPM, VaR; and CVaR. With the exception of the variance, these metrics can account for asymmetries as they can measure both left tail and right tail probabilities. This approach allows us to comprehensively examine hedging effectiveness of both short and long hedgers for both symmetric and asymmetric distributions and to see whether there is a dominant OHR estimation method that emerges across a broad range of hedging effectiveness metrics.

Our most important finding is that the presence of skewness in the return distribution reduces in-sample hedging effectiveness. Therefore hedges may fail during stressful market conditions when they are most required. We also find significant differences between the hedging effectiveness of short and long hedgers for each of the tail specific measures we apply, with the differences being more pronounced for the skewed distribution. This implies that hedgers who fail to use tail specific hedging performance metrics may chose inefficient hedging strategies that result in them being mishedged vis a vis their hedging objectives. Our results also show that the OLS model yields the best overall performance irrespective of the distributional characteristics of the index being hedged, or the hedging effectiveness criteria being applied. Based on this finding, it would appear that there is little to be gained in terms of hedging efficiency from the use of more complicated hedging strategies such as asymmetric GARCH models and that a simple hedging solution based on OLS is adequate for both normal and skewed distributions.

The remainder of the paper proceeds as follows. In section 2 we outline the models used for estimating optimal hedge ratios. In Section 3 we describe the metrics for measuring hedging effectiveness. Section 4 describes the data and summary statistics. Section 5 presents our empirical findings. Section 6 summarises and concludes.

**2.     Hedging Models**

The OHR is defined in the literature as the ratio that minimises the risk of the payoff of the hedged portfolio. The payoff of a hedged portfolio is given as:

$$+ r_s - \beta\ r_f \quad \text{(short hedger)} \tag{1a}$$

$$- r_s + \beta\ r_f \quad \text{(long hedger)} \tag{1b}$$

where $r_s$ and $r_f$ are the returns on the spot and futures respectively, and $\beta$ is the estimated OHR. We define a short (long) hedger as being long (short) the spot asset and short (long) the futures asset. In this study we utilise five different methods for estimating OHR's. The simplest models are a Hedge Ratio (HR) of zero (no hedge) and a 1:1 or Naïve hedge ratio where each unit of the crude oil contract is hedged with equivalent units of the futures contract. The third method we use is an OLS HR using the slope coefficient of a regression of the spot on the futures returns. An OHR estimated by OLS was first used by Ederington (1979) and has been applied extensively in the literature. Cecchetti et al., (1988) argue that the OLS method is not optimal because it assumes that the OHR is constant whereas time-varying volatility is the rule for financial time series and as the OHR depends on the conditional distribution of spot and futures returns, so too should the hedge ratio. We therefore use a rolling window OLS model to account for time varying effects. This is given as:

$$r_{st} = \alpha + \beta_t r_{ft} + \varepsilon_t \tag{2}$$

where $r_{st}$ and $r_{ft}$ are the spot and futures returns respectively for period t, $\varepsilon_t$ is the disturbance term and $\beta_t$ is the OHR. This can also be expressed as:

$$\beta_t = \frac{H_{sft}}{H_{ft}} \tag{3}$$

where $H_{ft}$ denotes the variance of futures returns and $H_{sft}$ is the covariance between spot and futures returns. We also use two additional estimation methods that allow the OHR to be time varying. These are a symmetric and an asymmetric GARCH model.

**The Symmetric Diagonal VECH GARCH Model (SDVECH)**

The first GARCH model that we use is the Diagonal Vech model proposed by Bollerslev, Engle and Wooldridge (1988). This model imposes a symmetric response on the variance and is useful for comparison of hedging estimation and performance as it has been extensively applied in the literature to generate OHR's (see, for example, Cotter and Hanly, 2006). This model is specified as follows:

$$r_{st} = \mu_{st} + \varepsilon_{st} \quad r_{ft} = \mu_{ft} + \varepsilon_{ft}, \quad \begin{bmatrix} \varepsilon_{st} \\ \varepsilon_{ft} \end{bmatrix} \Omega_{t-1} \sim N(0, H_t) \tag{4}$$

$$H_{st} = C_s + A_s \varepsilon_{st-1}^2 + B_s H_{st-1} \tag{5}$$

$$H_{ft} = C_f + A_f \varepsilon_{ft-1}^2 + B_f H_{ft-1} \tag{6}$$

$$H_{sft} = C_{sf} + A_{sf}\varepsilon_{st-1}\varepsilon_{ft-1} + B_{sf}H_{sf\,t-1} \qquad (7)$$

where $r_{st}, r_{ft}$ are the returns on spot and futures respectively, $\varepsilon_{st}, \varepsilon_{ft}$ are the residuals, $H_{st}, H_{ft}$ denotes the variance of spot and futures and $H_{sft}$ is the covariance between them. C is a 3x1 column vector, and A and B are 3x3 matrices. The matrices A and B are restricted to be diagonal. This means that the conditional variance of the spot returns depends only on past values of itself and past values of the squared innovations in the spot returns. The conditional variance of the futures returns and the conditional covariance between spot and futures returns have similar structures. Because of the diagonal restriction we use only the upper triangular portion of the variance covariance matrix, the model is therefore parsimonious, with only nine parameters in the conditional variance-covariance structure of the Diagonal VECH model to be estimated. This is subject to the requirement that the variance-covariance matrix is positive definite for all values of $\varepsilon_t$ in order to generate positive hedge ratios.

**The Asymmetric Diagonal VECH GARCH Model (ASDVECH)**

The second GARCH model that we use is an asymmetric extension of the SDVECH model. We require both a symmetric and an asymmetric GARCH model since we are examining their suitability for symmetric and asymmetric datasets. The key advantage of using an asymmetric model is that it is able to capture the asymmetries both within and between crude oil spot and futures markets. It therefore allows for different volatility response to negative and positive returns which means that dynamic hedging strategies will differ from those models that impose symmetry. The asymmetric GARCH model we use builds on the univariate GJR asymmetric GARCH model of Glosten et al, (1993). The GJR model allows the variance to respond differently to positive and negative return innovations through the use of an additional term designed to capture asymmetries. The multivariate model is similar to that used in de Goeij and Marquering (2004). It differs from the SDVECH model by changing the equations for the variances of the spot and futures given in equations (5) and (6) as follows:

$$H_{s_t} = C_s + A_s \varepsilon_{st-1}^2 + B_s H_{st-1} + D_s \varepsilon_{st-1}^2 I_{t-1} \quad (8)$$

$$H_{f_t} = C_f + A_f \varepsilon_{ft-1}^2 + B_f H_{ft-1} + D_f \varepsilon_{ft-1}^2 I_{t-1} \quad (9)$$

where $D$ is an additional term added to account for possible asymmetries, and $I$ is an indicator function whereby $I_{t-1} = 1$ if $\varepsilon_{st-1}, \varepsilon_{ft-1} < 0$, $I_{t-1} = 0$ if $\varepsilon_{st-1}, \varepsilon_{ft-1} > 0$. Therefore, the asymmetric effects are only applied to the variances involving lagged own residuals.

The advantage of GARCH models is that they can jointly estimate the conditional variances and covariances required for optimal hedge ratios and can also account for asymmetric effects in volatility when extended as appropriate (see, for example, Kroner and Sultan, 1993; De Goeij, et al, 2004). However, the performance of these models has been mixed. Over short time horizons and in-sample they have performed well (Conrad, Gultekin and Kaul, 1991), however, over longer hedging horizons and out-of-sample their performance has been poor (Lin et al, 1994). Brooks et al, (2002) use an asymmetric model to compare the hedging performance of symmetric and asymmetric GARCH. Their findings indicate that accounting for asymmetry may yield marginal improvements in hedging efficiency in-sample but will not improve out-of-sample hedging. There is no conclusive evidence that the GARCH class of models can consistently outperform the simpler Naïve and OLS based OHR's. However, the majority of studies have only applied very narrow performance criteria in evaluating hedging performance. In addition, the findings in general relate to financial asset hedging as relatively few studies have focused on commodity hedging as we do in this paper. We now turn to hedging effectiveness in the next section.

## 3. Hedging Effectiveness

We address a gap in the literature on commodity hedging by employing a number of hedging effectiveness metrics in addition to the variance. Each of these can distinguish between tail probabilities and can therefore measure performance for short and long hedgers for asymmetric distributions. The hedging effectiveness metrics we use are based on the Variance, LPM, VaR, and CVaR.

The performance metric generally applied in the literature on hedging is the percentage reduction in the variance of the spot (unhedged) position to the variance of the hedged portfolio. This was proposed by Ederington (1979) and has been widely adopted. It is given as:

$$HE_1 = 1 - \left[ \frac{VARIANCE_{HedgedPortfolio}}{VARIANCE_{UnhedgedPortfolio}} \right] \quad (10)$$

The use of the variance as a measure of risk has been criticised because it fails to distinguish between the tails of the distribution and therefore fails to differentiate in performance terms between short and long hedgers. Where distributions are asymmetric, the variance will over or underestimate risk. It is therefore not an adequate measure of risk for hedgers except where distributions are symmetric. Also, Lien (2005b) argues that the reason that the literature finds that the OLS hedging strategies tend to perform as well or better than more complex estimation models such as GARCH, can be attributed to the use of the Ederington effectiveness measure. Demirer et al (2005) argues that the variance performance criterion is only valid for OLS based hedging strategies and that using it to evaluate non-OLS OHR's will lead to the incorrect conclusion that OLS OHR's are the best from a hedging perspective.

Therefore, while the variance based measure is appropriate where the hedger seeks to minimise variance, in practice, hedgers may seek to minimise some measure of risk other than the variance. We therefore use three additional hedging effectiveness metrics that will allow us to make a robust comparison of different hedging models under both symmetric and asymmetric conditions.

The LPM is the first of a number of hedging effectiveness metrics that we use to address the shortcoming in the variance measure. The LPM distinguishes between the left and right tail of a distribution and can, therefore, measure risk in the presence of asymmetries as it is a function of the underlying distribution (Bawa, 1975). The LPM measures the probability of falling below a pre-specified target return. The LPM of order $n$ around $\tau$ is defined as

$$\text{LPM}_n(\tau;R) = E\{(\max[0, \tau - R])^n\} \equiv \int_{-\infty}^{\tau} (\tau - R)^n \, dF(R) \qquad (11)$$

where F(R) is the cumulative distribution function of the investment return R and $\tau$ is the target return parameter. The value of $\tau$ will depend on the level of return or loss that is acceptable to the investor. Some values of $\tau$ that may be considered are, zero or the risk free rate of interest. For a hedger a small negative return may be acceptable to reflect the cost of hedging. The parameter $n$ reflects the weighting applied to shortfalls from the target return. The more risk averse an investor the higher the weight ($n$) that would be attached. Fishburn (1977) shows that $0 < n < 1$ is suitable for a risk seeking investor, $n = 1$ for risk neutral, and $n > 1$ for a risk adverse investor. We can form a complete set of downside risk measures by changing the $\tau$ and $n$ parameters to reflect the position and risk preferences of different types of hedger. The more risk averse investors may set $\tau$ as the disaster level of return and have utilities that would reflect an LPM with $n = 2, 3....$ For example, an investor who views the consequences of a disaster level return as being unacceptable may use an order of LPM with $n = 5$.

Interest in one sided risk measures has increased in recent years as investors do not weight positive and negative outcomes evenly, but focus on downside or tail specific risk. The LPM therefore serves as an intuitive measure of risk that is in line with the risk preferences of many investors (Lee and Rao, 1988). Price and Nantell (1982) in a comparison of LPM and variance based measures of hedging effectiveness, find that they provide equivalent measures for normal distributions but that the LPM differentiates for distributions that are asymmetric as it is a tail specific measure that measures the left and right tail probabilities independently. A number of studies have compared the hedging effectiveness of short and long hedgers using the LPM methodology (see, for example, Demirer and Lien, 2003, and Demirer et al, 2005). The results from these studies indicate that hedging effectiveness for long hedgers differs from that of short hedgers with long hedgers deriving more benefit from hedging in terms of risk reduction as measured by reduced LPM's. Other studies that use the LPM include Lien and Tse (1998), who examined the consequences of time-varying conditional heteroskedasticity on optimal LPM based hedge ratios. They find that the optimal LPM hedge ratios are, on average, smaller than the minimum-variance hedge ratios. Lien and Tse (2000) argue that the minimum-variance hedge strategy is inappropriate for investors who are strongly risk averse such as those with utilities that reflect the higher orders of magnitude allowed for by the LPM. Since the majority of hedgers seek to limit losses, it may therefore, be appropriate for hedgers to seek to minimise a risk measure that is tail specific. Two additional metrics that are closely related to the LPM and which also have similar advantages as measures of risk for hedging are VaR and CVaR. Both of these measures are also tail specific and may be used to measure downside risk. These are considered in due course.

We calculate the lower partial moment using a target return of zero since the general aim of hedging is to avoid negative outcomes. We also use $n=3$ as the order of LPM, as this corresponds to a strongly risk averse investor. The metric we apply to evaluate hedging effectiveness based on the LPM is the percentage reduction in the LPM of each hedging estimation method as compared with a no hedge position[1].

This is calculated as:

$$HE_2 = 1 - \left[ \frac{LPM_{3HedgedPortfolio}}{LPM_{3UnhedgedPortfolio}} \right] \qquad (12)$$

The third hedging effectiveness metric is VaR[2]. This is the loss level of a portfolio over a certain period that will not be exceeded with a specified probability. VaR has two parameters, the time horizon (N) and the confidence level (x). Generally VaR is the $(100-x)^{th}$ percentile of the portfolio over the next N days. We calculate VaR using $x = 99$ and $N = 1$. This corresponds to the first percentile of the return distribution of each portfolio over a one-day period which is consistent with the OHR estimation period used in this study. Similar to the LPM, the performance metric employed is the percentage reduction in the VaR.

$$HE_3 = 1 - \left[ \frac{VaR_{1\%HedgedPortfolio}}{VaR_{1\%UnhedgedPortfolio}} \right] \qquad (13)$$

A shortcoming in VaR is that it is not a coherent measure of risk, as it is not subadditive. Also in practice two portfolios may have the same VaR but different potential losses. This is because VaR does not account for losses beyond the $(100-x)^{th}$ percentile. We address this by estimating an additional performance metric; Conditional Value at Risk (CVaR) that addresses the shortcoming in VaR as it is a coherent measure of risk.

Coherent measures of risk are subadditive. What this means is that the risk of two positions when added together is never greater than the sum of the risks of the two individual positions (see, for example, Artzner et al, 1999). Few studies in the hedging literature have applied VaR as a hedging effectiveness measure, however, Giot and Laurent (2003) use both the VAR and CVaR measures to examine the risk of short and long trading positions over a one day time horizon. They estimate volatility using both symmetric and asymmetric GARCH type approaches. Their findings show that symmetric models underperform models that account for asymmetry; however, their analysis is only applied to unhedged positions.

---

[1] The percentage reduction in the relevant performance measure is generally compared with a no-hedge position however there may be some cases where a no-hedge position does not yield the worst hedging performance, whereby the comparison is than made against the hedging model that yields the worst hedging performance.

CVaR is the expected loss conditional that we have exceeded the VaR. It is given as:

$$CVaR = E[L|L > VaR] \qquad (14)$$

This measures the expected value of our losses, L, if we get a loss in excess of the VaR. CVaR is preferable to the VaR because it estimates not only of the probability of a loss, but also the magnitude of a possible loss. In calculating CVaR we use the 1% confidence level which gives us the expected shortfall beyond the 1% VaR. The performance metric we use to evaluate hedging effectiveness is the percentage reduction in CVaR as compared with a no hedge position.

$$HE_4 = 1 - \left[ \frac{CVaR_{1\% \, HedgedPortfolio}}{CVaR_{1\% \, UnhedgedPortfolio}} \right] \qquad (15)$$

Next we turn to our application of both the hedging models and performance measures outlined to examine hedging effectiveness under conditions of asymmetry.

---

[2] As stated we choose a weight of n = 3 to describe a strongly risk averse investor. However, our measures of VaR and CVaR can also encompass attitudes to risk. VaR can be viewed as a special case of the LPM in equation (11) setting n = 0 and CVaR sets n = 1. By fixing the probability $LPM_0$, the corresponding VaR/CVaR can be calculated.

## 4. Data Description

Our data consist of daily commodity crude oil cash and futures data. Crude oil was chosen as it potentially exhibits significant asymmetric effects based on the reasons discussed earlier, because it is the largest traded commodity in the world in terms of monetary value, because of its economic importance and because it is widely hedged. It is therefore a suitable asset to use in an examination into the effects of asymmetry on hedging. The oil contract used is the NYMEX West Texas Light Sweet Crude contract which was chosen as it is used as an international oil pricing benchmark given its liquidity and price transparency. All data was obtained from Commodity Systems and daily returns were calculated as the differenced logarithmic prices. A continuous series was formed with the contract being rolled over by volume. This means that the price of the largest traded contract by volume was used and that the price switched from one contract to the next when that contract's volume fell below the volume of the next traded contract.

Since we are examining the influence of asymmetry and its effects on hedging effectiveness, we have used an innovative approach in choosing the data. The crude oil contract and the sample period were chosen on the basis of their distributional characteristics as we required data with both symmetric and asymmetric distributions. A single crude oil contract was used for the sample period from 1 January 2000 – 31 December 2001. We broke down the data into two separate periods, one symmetric and one asymmetric. The first dataset corresponds to the period 1 January 2000 – 31 December 2000 and is termed Period 1. The second dataset corresponds to the period from 1 January 2001 – 31 December 2001 and is termed Period 2.

For each period, the first 160 observations comprise the in-sample period. The remaining 100 observations were used to facilitate out-of-sample comparisons. Summary statistics for the data are displayed in Table I. The characteristics of the return distributions for both series are markedly different. From table 1, we can see that period 1 is non-symmetric and is characterised by negative skewness. In contrast, Period 2 can be characterised as Gaussian given the insignificant skewness figure and our failure to reject the hypothesis of normality based on the Bera-Jarque statistic. The use of a symmetric period also allows us to compare the hedging effectiveness under Gaussian market conditions of each of the different models using the hedging effectiveness metrics previously outlined.

**Table 1:      Summary Statistics of Daily Crude Oil Spot and Futures Returns**

| Sample Period | PERIOD 1 Crude oil-Asymmetric 2001 | | PERIOD 2 Crude oil-Symmetric 2002 | |
|---|---|---|---|---|
| **Index** | Spot | Futures | Spot | Futures |
| **Mean** | -0.0012 | -0.0012 | 0.0018 | 0.0017 |
| **Min** | -0.1653 | -0.1654 | -0.0586 | -0.0627 |
| **Max** | 0.1039 | 0.0807 | 0.0613 | 0.0595 |
| **Std Dev** | 0.0277 | 0.0261 | 0.0212 | 0.0208 |
| **Skewness** | -0.633* | -1.003* | -0.066 | -0.041 |
| **Kurtosis** | 6.746* | 6.762* | 0.421 | 0.384 |
| **B-J** | 512.36* | 541.14* | 2.112 | 1.678 |
| **LM** | 7.17 | 1.53 | 9.48 | 11.41** |
| **Stationarity** | -16.25* | -15.43* | -16.83* | -17.51* |

Notes:     Summary statistics are presented for the log returns of each spot and futures series.  The skewness statistic measures asymmetry where zero would indicate a symmetric distribution. The kurtosis statistic measures the shape of a distribution as compared with a normal distribution. Figures reported for kurtosis are for excess kurtosis where a value of zero would indicate a normal distribution.  The Bera-Jarque (B-J) statistic combines skewness and kurtosis to measure normality.  LM is the Lagrange Multiplier test proposed by Engle (1982). The test statistic for B-J, LM tests are distributed $\chi^2$. Stationarity is tested using the Dickey-Fuller unit root test.  *Denotes Significance at the 1% level. **Denotes Significance at the 5% level.

The data were checked for stationarity using Dickey Fuller unit root tests. As expected, the log returns for both series are stationary. Stationarity is important as a non-stationary series may lead to spurious regressions and invalidate the estimated optimal hedge ratios. LM tests were used to check for heteroskedasticity with only the futures returns for Period 2 showing significant GARCH effects. This may limit the theoretical advantages of the GARCH models over the hedging strategies that assume a constant variance.

## 5. Empirical Findings

We compare the short and long hedging effectiveness of five different OHR estimation methods for distributions with symmetric and asymmetric characteristics. We construct hedge portfolios using (1a) and (1b) and then evaluate hedging effectiveness over a 1-day re-balancing period using the performance metrics outlined in section 3. The results of the hedging estimation models are quite standard and are therefore not reproduced in detail[3]. Figure 1 presents OHR's for the Rolling OLS, symmetric and asymmetric GARCH models.

**Figure 1:** Optimal Hedge Ratios

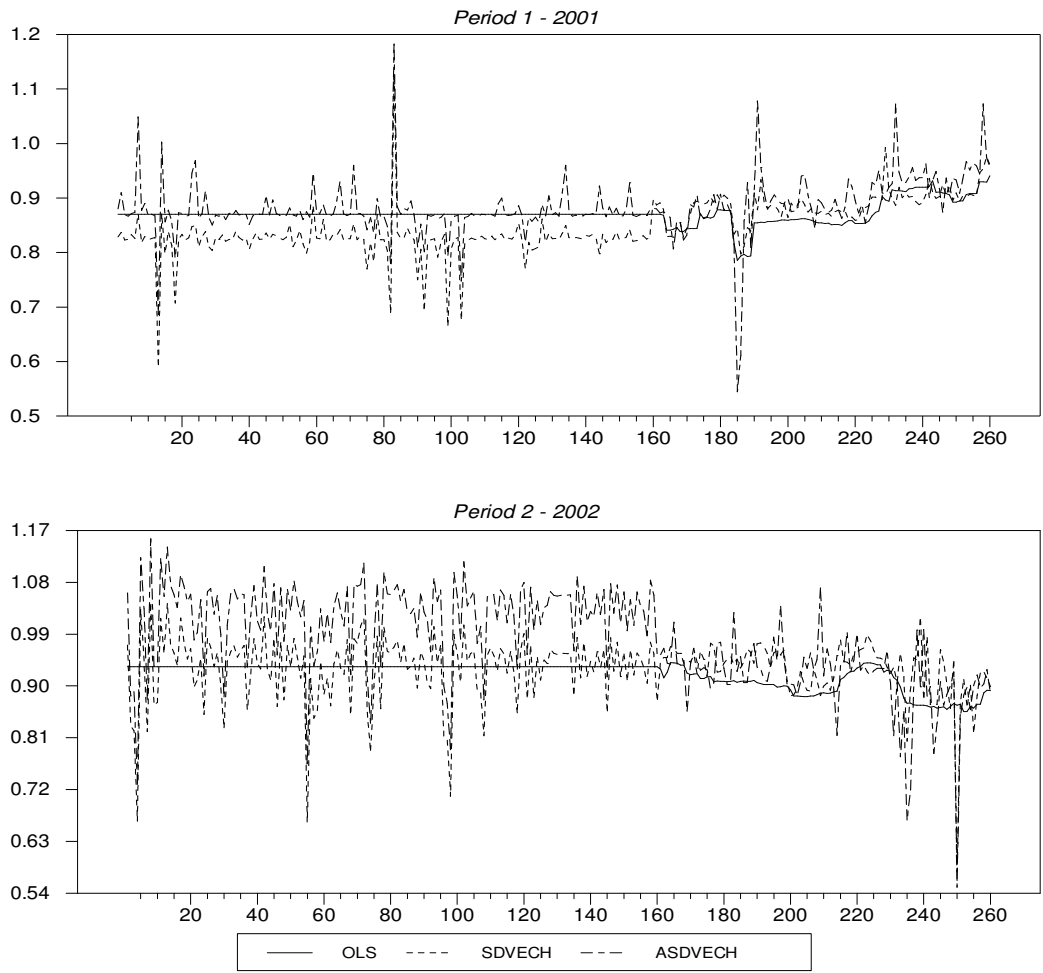

Notes: Figure 1 displays OHR's for both asymmetric (period 1) and symmetric (period 2) distributions.

---

[3] Full details are available on request

Figure 1 reveals significant differences between the OHR's for the different models for both periods. The gap between the GARCH model OHR's for the asymmetric period is to be expected however there is also a large difference between the OHR's for the GARCH models for the symmetric period. This indicates that although the Period 2 dataset is symmetric as measured by skewness, there may still be an asymmetric response to positive and negative return innovations that is being picked up by the asymmetric GARCH model and this is considered in due course. The OHR's also appear to be stationary, therefore, while we would expect to see different model hedging performance for each of the different models, the performance gap may be narrow in many cases.

Table 2 presents results for hedging effectiveness. Comparisons are made between the hedging effectiveness of short and long hedgers for both the asymmetric data (Period 1) and symmetric data (Period 2). Statistical comparisons are made using Efrons (1979) bootstrap methodology by employing t-tests of the differences between short and long hedgers based on the point estimates of our results[4]. This approach is also adopted in tests of model hedging effectiveness and allows us to make statistical as well as economic inferences from our results. Comparisons are carried out both in-sample and out-of-sample and across each of the hedging effectiveness metrics.

Examining first general hedging performance, we can observe large differences in hedging effectiveness for the asymmetric and symmetric datasets. General hedging performance across all models and risk measures is worse for the asymmetric period. For example, variance reductions for the symmetric period are of the order of 85% as compared with an average of 64% in-sample for the asymmetric period when all hedging models are used. This finding persists if we use the tail specific measures with average differences in hedging performance between the symmetric and asymmetric periods of 27% and 31% for short and long hedgers respectively. We derive a number of conclusions from this result. First, the in-sample hedging performance is better for symmetric distributions suggesting that hedging oil commodity using futures is of limited use during volatile periods where returns may be skewed. Secondly, hedging effectiveness for oil commodities in general is not as good as for other assets such as equity indices. This may be attributed to the relationship between spot and futures prices in commodities markets which may be volatile depending on changes in the underlying supply and demand conditions. An interesting result is that hedging is not as effective in reducing VaR and CVaR as it is at reducing the LPM, with differences in hedging effectiveness typically ranging from 15%-50%. We can also see that the different hedging models yield poor performance in terms of reducing the VaR and especially the CVaR. Using the best performing models for example, the in-sample results for the skewed period 1 series indicate only an 11% reduction in the CVaR for short hedgers and a 29% reduction for long hedgers.

These results may be related to the ability of the VaR and CVaR metrics to model extreme tail events whereas the LPM is a more general metric that doesn't pick up the most extreme outliers in the same way[5]. This result indicates that hedging oil commodities may be quite limited in reducing extreme losses as measured by the VaR or CVaR. These hedging effectiveness metrics may therefore be appropriate for use by strongly risk averse investors as a measure of risk for distributions that may be skewed by a small number of observations.

---

[4] The returns of the hedged portfolios as compiled using equations 1a and 1b were bootstrapped by resampling with replacement from the returns. 100 simulations were used allowing for the construction of confidence intervals around each point estimate.

[5] While each of these measures is one sided, the LPM with a target rate set t=0 will include all observations less than 0, whereas both the VaR and CVaR calculated at the 1% interval will include only extreme observations located in the left tail of the distribution. Also, modelling the tail of the distribution is more statistically reliable as compared with the exceeding method that relates to the LPM.

**Table 2:     Hedging Effectiveness – Short Vs Long Comparison**

| (1) | (2) HE$_1$ Variance (x10$^{-2}$) | (3) HE$_2$ LPM (x10$^{-2}$) | (4) HE$_3$ VaR (x10$^{-2}$) | (5) HE$_4$ CVaR (x10$^{-2}$) | (6) HE$_1$ Variance (x10$^{-2}$) | (7) HE$_2$ LPM (x10$^{-2}$) | (8) HE$_3$ VaR (x10$^{-2}$) | (9) HE$_4$ CVaR (x10$^{-2}$) |
|---|---|---|---|---|---|---|---|---|
| | **Panel A: SHORT HEDGERS** | | | | **Panel B: LONG HEDGERS** | | | |
| | | | | **IN SAMPLE** | | | | |
| **PERIOD 1** **CRUDE OIL-Asymmetric** | | | | | | | | |
| None | 00.00 | 00.00 | 00.00 | **11.61*** | 00.00 | 00.00 | 00.00 | 00.00 |
| Naïve | 64.45 | **76.18*** | 13.19 | 0.00 | 64.45 | 36.55 | **26.38*** | **29.07*** |
| OLS | 65.90 | **69.97*** | 20.61 | 9.91 | 65.90 | 55.27 | **26.55*** | **22.57*** |
| SDVECH | 64.24 | **72.13*** | 18.25 | 3.97 | 64.24 | 44.93 | **30.05*** | **24.53*** |
| ASDVECH | 65.07 | **74.17*** | 17.77 | 3.05 | 65.07 | 43.30 | **29.25*** | **26.09*** |
| **PERIOD 2** **CRUDE OIL-Symmetric** | | | | | | | | |
| None | 00.00 | 00.00 | 00.00 | 00.00 | 00.00 | 00.00 | 00.00 | 00.00 |
| Naïve | 82.95 | **92.02** | 42.01 | 33.18 | 82.95 | 87.05 | **47.17** | **51.43*** |
| OLS | 83.33 | 90.08 | 45.59 | **48.94*** | 83.33 | 90.82 | 47.77 | 40.47 |
| SDVECH | 84.48 | **93.29** | 45.45 | 36.26 | 84.48 | 89.26 | **55.38*** | **56.17*** |
| ASDVECH | 83.52 | **92.76** | 43.97 | 33.86 | 83.52 | 88.00 | **52.74*** | **54.35*** |
| | | | | **OUT OF SAMPLE** | | | | |
| **PERIOD 1** **CRUDE OIL-Asymmetric** | | | | | | | | |
| None | 00.00 | 00.00 | 00.00 | 00.00 | 00.00 | 00.00 | 00.00 | 00.00 |
| Naïve | 80.54 | 79.97 | **64.02*** | **59.77*** | 80.54 | **93.56*** | 40.77 | 36.06 |
| OLS | 81.11 | 85.37 | 51.69 | **62.99*** | 81.11 | **93.53*** | 52.08 | 47.85 |
| SDVECH | 80.22 | 80.87 | **62.88*** | **64.23*** | 80.22 | **94.12*** | 37.44 | 40.00 |
| ASDVECH | 80.37 | 81.62 | **62.86*** | **64.90*** | 80.37 | **94.14*** | 37.58 | 40.55 |
| **PERIOD 2** **CRUDE OIL-Symmetric** | | | | | | | | |
| None | 00.00 | 00.00 | 00.00 | **2.90*** | 00.00 | 00.00 | 00.00 | 00.00 |
| Naïve | 70.38 | **76.82*** | 15.40 | 0.00 | 70.38 | 62.83 | **33.43*** | **26.25*** |
| OLS | 71.55 | **75.63** | 24.40 | 16.75 | 71.55 | 71.64 | **29.94** | 19.71 |
| SDVECH | 71.55 | **78.25*** | 17.58 | 3.06 | 71.55 | 66.62 | **35.28*** | **28.25*** |
| ASDVECH | 72.43 | **81.02*** | 19.05 | 2.84 | 72.43 | 66.92 | **35.50*** | **36.48*** |

Notes:   Figures are in percentages. HE$_1$ – HE$_4$ give the percentage reduction in the performance metric from the hedged model as compared with the worst hedge position. For example, short hedging the crude oil contract for Period 1 with the NAIVE model yields a 76.18% in-sample reduction in the variance as compared with a No-Hedge strategy. Comparisons are made between the performance of short and long hedgers on a metric by metric basis. The highlighted figures indicate where one set of hedgers outperform the other for a given hedging effectiveness metric. * indicates significant difference at the 1% confidence level.

A key point is that hedges may not be as effective in volatile markets that are skewed. In hedging terms, this means that investors may face the risk that their hedges will not fulfil their function of risk reduction during stressful markets conditions when they are most needed. When we examine the out-of-sample results, we find that only the variance metric supports the in-sample results and that the tail specific metrics tend to show better hedging effectiveness for the asymmetric period as compared with the symmetric. However, this result is economically significant for the short hedgers only and may be biased given the relatively small sample. Also, again we can see the limited use of hedging in reducing risk as measured by VaR and CVaR metrics. This is an indication that extreme values may persist for commodity oil returns that cannot be hedged away.

The second comparison we make is between the hedging performance of short and long hedgers for asymmetric as compared to symmetric periods[6]. For Period 1 (Asymmetric) the hedging effectiveness differences between short and long hedgers are significant at the 1% level in every single case in-sample and in 92% of cases out-of-sample whereas for Period 2 (Symmetric) they are significant in only 50% of cases in-sample and 69% of cases out-of-sample. The differences between short and long hedgers even for the symmetric period indicate that one must be careful of using the skewness statistic to measure asymmetry as positive and negative skewness may cancel each other out. It also indicates the importance of using tail specific hedging effectiveness metrics irrespective of the characteristics of the return distribution. This finding is supported across each of the different tail specific metrics of hedging performance.

Table 3 presents absolute figures for the performance effectiveness measures of each of the hedging models. We compare the results for Period 1 and Period 2 to see whether the distributional characteristics will have an effect on the hedging effectiveness of different models. In-sample, the best overall hedging model for the asymmetric Period 1 is the OLS model in 50% of cases. This result applies to both short and long hedgers. The best overall model for the symmetric Period 2 is the SDVECH model. Out-of sample results indicate that the OLS and

---

[6] For each hedging model we compare the performance metric for short hedgers with the performance metric for long hedgers again employing bootstrap confidence intervals. Taking the asymmetric Period 1 for example, the difference between the Naïve model VaR figures of 13.19 and 26.38 respectively is statistically significant at the 1% level.

**Table 3:** **Statistical Comparison of Hedging Model Performance**

| (1) | (2) Variance (x10⁻³) | (3) LPM (x10⁻⁵) | (4) VaR (x10⁻²) | (5) CVaR (x10⁻²) | (6) Variance (x10⁻³) | (7) LPM (x10⁻⁵) | (8) VaR (x10⁻²) | (9) CVaR (x10⁻²) |
|---|---|---|---|---|---|---|---|---|
| | **Panel A: SHORT HEDGERS** | | | | **Panel B: LONG HEDGERS** | | | |
| | | | | **IN-SAMPLE** | | | | |
| **PERIOD 1** | | | | | | | | |
| **CRUDE OIL-Asymmetric** | | | | | | | | |
| NONE | 0.481* | 1.375* | 5.278* | 5.700 | 0.481* | 0.722* | 5.917* | 8.107* |
| NAIVE | 0.171 | 0.328 | 4.582* | 6.449* | 0.171 | 0.458 | 4.356 | 5.750 |
| OLS | 0.164 | 0.413* | 4.190 | 5.810 | 0.164 | 0.323 | 4.346 | 6.277* |
| SDVECH | 0.172 | 0.383 | 4.315 | 6.193 | 0.172 | 0.398* | 4.139 | 6.118* |
| ASDVECH | 0.168 | 0.355 | 4.340 | 6.252 | 0.168 | 0.410* | 4.186 | 5.992* |
| | | | | | | | | |
| **PERIOD 2** | | | | | | | | |
| **CRUDE OIL-Symmetric** | | | | | | | | |
| NONE | 0.522* | 1.075* | 5.115* | 5.587* | 0.522* | 0.895* | 5.518* | 6.131* |
| NAIVE | 0.089* | 0.086* | 2.966* | 3.733* | 0.089* | 0.116* | 2.915* | 2.978* |
| OLS | 0.087* | 0.107* | 2.783 | 2.853 | 0.087* | 0.082 | 2.882* | 3.650* |
| SDVECH | 0.081 | 0.072 | 2.790 | 3.561* | 0.081 | 0.096* | 2.462 | 2.687 |
| ASDVECH | 0.086 | 0.078 | 2.866 | 3.695* | 0.086 | 0.107* | 2.608* | 2.799* |
| | | | | **OUT-OF-SAMPLE** | | | | |
| **PERIOD 1** | | | | | | | | |
| **CRUDE OIL-Asymmetric** | | | | | | | | |
| NONE | 1.240* | 2.862* | 12.183* | 16.531* | 1.240* | 8.110* | 9.399* | 10.395* |
| NAIVE | 0.241 | 0.573* | 4.383 | 6.650* | 0.241 | 0.522 | 5.567* | 6.647* |
| OLS | 0.234 | 0.419 | 5.885* | 6.118 | 0.234 | 0.524 | 4.504 | 5.421 |
| SDVECH | 0.245 | 0.548* | 4.522 | 5.913 | 0.245 | 0.476 | 5.880* | 6.237* |
| ASDVECH | 0.243 | 0.526* | 4.525 | 5.803 | 0.243 | 0.475 | 5.867* | 6.180* |
| | | | | | | | | |
| **PERIOD 2** | | | | | | | | |
| **CRUDE OIL-Symmetric** | | | | | | | | |
| NONE | 0.341* | 0.634* | 4.278* | 4.312* | 0.341* | 0.469* | 4.974* | 5.353* |
| Naïve | 0.101* | 0.147* | 3.619* | 4.441* | 0.101* | 0.174* | 3.311 | 3.948* |
| OLS | 0.097 | 0.154* | 3.234 | 3.697 | 0.097 | 0.133 | 3.485* | 4.298* |
| SDVECH | 0.097 | 0.138* | 3.526* | 4.305* | 0.097 | 0.157* | 3.219 | 3.841* |
| ASDVECH | 0.094 | 0.120 | 3.463* | 4.315* | 0.094 | 0.155* | 3.208 | 3.400 |

Notes: Statistical comparisons are made for each hedging model against the best performing model. For example, short hedging period 1, if we examine column 1, we can see that the in-sample hedging effectiveness of the no hedge model is significantly different than the best performing OLS model at the 1% level. * Denotes significance at the 1% level.

**Table 4:** **Summary of Best Performing Hedging Model**

| (1) | (2) Variance | (3) LPM | (4) VaR | (5) CVaR | (6) Variance | (7) LPM | (8) VaR | (9) CVaR |
|---|---|---|---|---|---|---|---|---|
| | **Panel A: SHORT HEDGERS** | | | | **Panel B: LONG HEDGERS** | | | |
| | | | | **IN-SAMPLE** | | | | |
| **PERIOD1** Asymmetric | OLS | NAÏVE | OLS | NONE | OLS | OLS | SDVECH | NAÏVE |
| **PERIOD 2** Symmetric | SDVECH | SDVECH | OLS | OLS | SDVECH | OLS | SDVECH | SDVECH |
| | | | | **OUT-OF-SAMPLE** | | | | |
| **PERIOD1** Asymmetric | OLS | OLS | NAÏVE | ASDVECH | OLS | ASDVECH | OLS | OLS |
| **PERIOD 2** Symmetric | ASDVECH | ASDVECH | OLS | OLS | ASDVECH | OLS | ASDVECH | ASDVECH |

Notes: Table 4 provides a summary of the best hedging model for both short and long hedgers for each measure of hedging effectiveness. For example the best hedging model in terms of risk reduction for Period 1 in-sample is the OLS model in four out of eight cases for both sets of hedgers.

ASDVECH models are the best overall performers for periods 1 and 2 respectively. While this demonstrates that the OLS and GARCH models tend to provide better hedging performance than a naïve hedge, a key issue is whether there are significant differences in the performance of the different hedging models. To test this we compared the performance of the different hedging models again employing Efrons (1979) bootstrap methodology. We find significant differences between model hedging performances in almost 70% of cases both in-sample and out-of-sample however if we exclude the no-hedge model, this figure drops by around 10%. Looking just at the OLS and GARCH models, we can see that there is little to choose between them as the absolute differences between models are small and not economically significant. This may in part be attributable to the low incidence of heteroskedasticity of our sample.

Table 4 provides a summary of the best performing models. Examining both in-sample and out-of-sample results, we can see that for an asymmetric distribution the OLS model is clearly the best performer across all hedging effectiveness metrics. For a symmetric distribution the OLS model again performs strongly although the GARCH models perform best in a small majority of cases. This is clear evidence of the failure of the asymmetric GARCH model to provide better hedging performance for skewed distributions. The failure of the asymmetric GARCH model to provide better performance for asymmetric distributions contrasts with that of Giot and Laurent (2003). This may relate to the earlier point relating to the ability of the ASDVECH model to model positive and negative return innovations separately, however, this is not the same as being able to model skewness in the distribution. Taken together with the results from table 3, we would have to conclude that there is little to be gained from the more complex GARCH models in performance terms over the simpler OLS model irrespective of the characteristics of the return distribution. This finding supports the broad literature on optimal hedging and indicates that the OLS model provides an efficient outcome across a selection of risk measures. It is therefore in support of the criticism by Lien (2005b) who suggests that the failure of more complicated hedging estimation models to outperform OLS are a result of the widespread use of the variance reduction criterion.

## 6. Summary and Conclusions

This paper compares the hedging effectiveness of commodity oil futures for both symmetric and asymmetric distributions. We also compare the hedging effectiveness of short and long hedgers using a variety of hedging estimation methods. A gap in the literature on optimal hedging is addressed by applying a comprehensive set of hedging effectiveness measures that are tail specific. We find that in-sample hedging effectiveness is significantly reduced by the presence of skewness in the return distribution. This is an important finding as it indicates that hedging may not be as effective during volatile periods and therefore investors may not be effectively hedging during the periods when they most require it.

We also find larger differences in hedging performance between the short and long hedgers for the asymmetric distribution when compared with a symmetric distribution. Therefore the use of one-sided hedging performance measures that are consistent with modern risk management techniques such as VaR and CVaR is to be recommended, as the traditional variance reduction criterion is not adequate and will provide inaccurate measures of risk for different types of hedgers. Also, we find that the best hedging estimation model to emerge is the OLS model. This provides the best overall hedging performance across all measures of hedging effectiveness for both short and long hedgers. While the GARCH models perform well for the symmetric distribution, the differences in performance as compared with the OLS model are not significant. We also find that employing a GARCH model that allows for asymmetries yields no significant improvement in performance. This finding suggests that there is little economic benefit to be gained by the use of more complex hedging estimation models over the simpler OLS model irrespective of the characteristics of the return distribution.